\documentclass[11pt]{article}
\usepackage{hyperref}
\pdfoutput=1
\begin{document}
\title{Inviscid Simulations of Interacting Flags \\ and Falling Sheets}
\author{Silas Alben \\
\\\vspace{6pt} School of Mathematics \\ Georgia Institute of
Technology, Atlanta, GA 30332-0160, USA}
\maketitle
\begin{abstract}
We present a fluid dynamics video showing simulations of flexible bodies flapping and falling in an inviscid fluid. Vortex sheets are shed from the trailing edges of the bodies
according to the Kutta condition. For interacting flags, a sampling of synchronous and asynchronous
states are shown. For falling flexible sheets, the basic behavior is a repeated series
of accelerations to a critical speed at which the sheet buckles, and rapidly
decelerates, shedding large vortices. Examples of persistent circling,
quasi-periodic flapping, and more complex trajectories are shown.
\end{abstract}

The video is shown in
\href{http://ecommons.library.cornell.edu/bitstream/1813/13743/2/TwoFlagsAndFallingMpeg2.mpg}{high-resolution} and
\href{http://ecommons.library.cornell.edu/bitstream/1813/13743/3/TwoFlagsAndFallingMpeg1.mpg}{low-resolution} files.
The first half of the video shows interacting flags, and the second half shows falling
flexible sheets.

\section{Simulations of Interacting Flags}

The video begins with ``{\bf Example of asynchronous dynamics for side-by-side
flags}.'' The flags are green solid lines, and
the vortex sheets they shed are blue dotted lines. The leading edges
are spaced apart by a distance equal to 60\% of the flags' lengths.
At this close distance, the flags do not synchronize. Their ends
interact strongly, which destabilizes any trend toward synchrony.
The flags do flap nearly out of phase for sustained periods, and
when the separation distance is increased to 90\% of the flags'
lengths, they flap synchronously and with opposite phase. Further
increases in distance yield a monotonic variation of phase, with the
flags flapping in phase at a separation of 2.5 flag lengths.

The next segment is ``{\bf Example of synchronous dynamics for tandem
flags}.'' The leader flag is green and the follower flag is orange.
The follower intercepts the vortex wake of the leader, which passes along
the follower tangentially, and influences its vortex shedding. The two wakes
merge, and the additional stimulation from the leader wake causes the
follower to flap with an amplitude and drag which are larger than that for the
leader.

The spacing between the leader and follower is increased in the subsequent clip,
``{\bf Example of asynchronous dynamics for tandem flags}.''  This
spacing does not allow the same-signed vorticity
to coincide when the flags' wakes meet. The follower flaps irregularly, and with
an amplitude and drag which are smaller on average than in the synchronous case,
and close to that of the leader. The black stars show point vortices,
which approximate the vortex sheets downstream for computational efficiency.

\section{Simulations of Falling Sheets}

The second topic of the video is introduced by the clip entitled ``{\bf Falling flexible sheet
trajectories: Example of buckling while falling}.'' The moving solid orange line
is a flexible fiber, falling under gravity, and shedding a vortex sheet (blue line) from its
trailing edge. The two control parameters are the sheet density normalized by fluid
density ($R_1$, 0.3 here) and the sheet rigidity normalized by fluid inertia ($R_2$, 2.4 here).
The basic behavior is a repeated series of accelerations to a critical speed at which the sheet buckles, and rapidly decelerates, shedding large vortices. The still frames which
surround the moving picture give sample trajectories for many different initial falling angles
and different sheet
rigidities. These paths show a diversity of punctuated falling and circling behaviors (circling is seen for
$R_2$ equal to 10 and above). The still frame labeled ``$R_1=0.3$, $R_2=2.4$'' shows in light blue
the trajectory traced by the orange fiber as it falls.

The final clip, ``{\bf Falling flexible sheet
trajectories: Examples of quasi-periodic flapping},'' shows an alternative falling behavior.
For a range of smaller $R_1$ and $R_2$ (two examples are shown), the body flaps steadily as
it falls. The example to the left is an asymmetric flapping state. The example to the right
shows symmetric flapping with a simple period. The drag encountered
by these flapping bodies balances the acceleration from gravity. The still panels again
show examples of different falling trajectories as parameters are varied. The blue trajectories
in the still frame labeled ``$R_1=0.3$, $R_2=1$''correspond to states of flapping while falling,
encountered for many different initial falling angles at these parameter values.

\end{document}